\RequirePackage{silence}
\documentclass[fleqn,usenatbib,useAMS]{mnras} 
\usepackage[utf8]{inputenc}
\DeclareUnicodeCharacter{2212}{-}
\usepackage{graphicx}
\usepackage{threeparttable}
\usepackage{scrextend}
\usepackage{tablefootnote}
\usepackage{amsmath}

\usepackage{amsfonts}
\usepackage{float}
\usepackage{bm}
\usepackage{ae,aecompl}
\usepackage{array}
\usepackage{soul}
\usepackage{mathtools}
\usepackage{multirow}
\usepackage{bigdelim}
\newcolumntype{_}{>{\global\let\currentrowstyle\relax}}
\newcolumntype{^}{>{\currentrowstyle}}

\newcommand{\appropto}{\mathrel{\vcenter{
			\offinterlineskip\halign{\hfil$##$\cr 
				\propto\cr\noalign{\kern2pt}\sim\cr\noalign{\kern-2pt}}}}}

\title[The origin of Ear-like Structure in SNRs]{Particle Acceleration along Magnetic Fields as the Origin of Ear-like Structures in Supernova Remnants} 

\author[Yu and Fang]
{Huan Yu$^{1}$, Jun Fang$^{2}$\thanks{
E-mail:  \mbox{fangjun@ynu.edu.cn}}\\
$^{1}$Department of Physical Science and Technology, Kunming University, Kunming, China\\
$^{2}$Department of Astronomy, Key Laboratory of Astroparticle Physics of Yunnan Province, Yunnan University, Kunming, China\\
}

\pubyear{2025}
\begin{document}
\label{firstpage}
\pagerange{\pageref{firstpage}--\pageref{lastpage}}

\maketitle

\begin{abstract}
The origin of ear-like structures with two opposite lobes extending from the shell of supernova remnants (SNRs) remains a subject of active debate, with proposed mechanisms ranging from jet-driven explosions to interactions with bipolar circumstellar media. We present a novel investigation using Particle-in-Cell (PIC) simulations to examine the dynamics of a spherical ejecta in magnetized media, revealing a different formation mechanism for these structures. Our results demonstrate that particle acceleration is significantly enhanced along the magnetic-field direction, producing substantial modifications to shock morphology. These effects naturally generate elongated protrusions bearing a remarkable similarity to the observed SNR ear structures.  This magnetic alignment scenario represents a distinct mechanism from the existing jet-based and circumstellar interaction models. Our findings suggest that magnetic field orientation may serve as a crucial diagnostic for distinguishing between different ear formation mechanisms in SNRs.

\end{abstract}


\section{Introduction}\label{sec1}

The morphological characteristics of SNRs provide crucial insights into both the explosion mechanisms of their progenitors and the properties of the surrounding circumstellar medium (CSM). Among various morphological peculiarities, the presence of protrusions or "ears" has been observed in a significant fraction of SNRs \citep{2017MNRAS.472.1770B}. These antisymmetric structures, typically appearing as two opposite lobes extending from the main shell, have been detected across different SNR types and evolutionary stages, suggesting they may originate from fundamental processes in supernova (SN) explosions or their immediate aftermath.

Two principal mechanisms have been proposed to explain the formation of these ear-like structures. The first, advocated by \citet{2017MNRAS.468.1226G}, attributes the formation of ears to energetic jets launched during or shortly after the core collapse explosion. Through the analysis of a sample of core collapse SNRs (CCSNRs), they found that approximately 29-43\% exhibit clear ear features. Under simple geometrical assumptions, they estimated that the kinetic energy required to inflate these ears ranges from 5-15\% of the total explosion energy, supporting the jet feedback mechanism (JFM) as a viable explosion mechanism for massive stars.


An alternative scenario attributes the formation of ears to the interaction of SN ejecta with a bipolar circumstellar medium (CSM). \citet{2021MNRAS.502..176C} demonstrated through hydrodynamic simulations that when the forward shock breaks out of an equatorially confined CSM, it produces long-lived, opposite protrusions, naturally explaining observed features like shocked CSM accumulation and the size variation of ears. Specifically within the context of Type Ia SNe, \citet{2020Galax...8...38C} proposed that such a bipolar CSM originates from a planetary nebula (PN) preceding the explosion. Their simulations showed that the delay time between the PN and SN phases dictates the CSM's structure and that this model successfully reproduces the key characteristics of Kepler's SNR, including its ears, thereby offering a progenitor model without a surviving donor star.

Although both mechanisms can produce morphologically similar structures, they differ in their predicted orientation and formation timescales. The jet model typically produces ears aligned with the presumed jet axis (often polar), whereas the bipolar CSM interaction forms ears preferentially at the equatorial plane. Additionally, the jet mechanism could operate at various stages of the evolution of the SNR, while the CSM interaction scenario specifically links ear formation to the early shock breakout phase. To observationally distinguish between these mechanisms, detailed studies of specific SNRs like the bilateral remnant SN 1006 are crucial.

The bilateral supernova remnants are supernova remnants with axisymmetric structures and two bright arcs, serving as important objects for studying the magnetic field structure around SNRs \citep{1998ApJ...493..781G}. SN 1006 is a prototypical bilateral supernova remnant characterized by a shell structure with two bright limbs in the northeast (NE) and southwest (SW) directions, and prominent protrusion structures are also observed at intermediate positions along these limbs \citep{1986AJ.....92.1138R,2003ApJ...589..827B,2004A&A...425..121R,Miceli2009,2014ApJ...781...65W,2015MNRAS.453.3953L}.
\citet{2013AJ....145..104R} investigates the radio polarization signature of SN 1006, a bilateral supernova remnant, to explore the relationship between magnetic field orientation, particle acceleration efficiency, and remnant morphology. In the bright lobes (polar caps), the quasi-parallel shock geometry facilitates efficient particle acceleration and magnetic field amplification, generating turbulence. Conversely, in the SE rim, the quasi-perpendicular shock geometry results in inefficient acceleration and preserves the ordered ambient field. The study concludes that the ambient magnetic field is aligned nearly parallel to the Galactic plane, supporting a polar cap model for the bilateral morphology.

The peculiar morphology of the Type Ia SNR SN 1006 has been investigated by numerical simulations by \cite{2020MNRAS.491.2460F}. Their 3D hydrodynamical simulations demonstrated that the observed bilateral structure with protrusions in the northeast and southwest limbs, along with a northwest filament, can be explained by the SNR expanding into an ambient medium with a density discontinuity. The model shows that the hemisphere expanding into a denser medium develops a smaller radius, and specific lines of sight intersecting these asymmetric structures reproduce the observed protrusions and filament. This work provides important insights into how density variations in the interstellar medium can shape SNR morphologies, though it does not account for all observed asymmetries or include particle acceleration effects.

SN 1006 serves as a key laboratory for investigating shock modification due to efficient particle acceleration. \cite{Miceli2009} conducted a comprehensive spatially resolved spectral analysis of \textit{XMM-Newton} observations of SN 1006, revealing that the thermal X-ray emission is predominantly associated with shocked ejecta, while the non-thermal synchrotron emission dominates in the bright limbs. Their results indicate that the efficient acceleration of particles modifies the shock structure throughout the entire rim, leading to a reduced distance between the blast wave and the contact discontinuity compared to unmodified shock models. These findings highlight the profound impact of particle acceleration on the morphology and thermal properties of SNRs, providing critical insights into the efficiency of cosmic-ray production and the dynamics of shock evolution.

We used particle-in-cell (PIC) numerical simulations to investigate the dynamical processes of high-velocity ejecta propagating into uniform magnetized media. Our simulations reveal significantly enhanced particle acceleration along the magnetic field direction, which profoundly modifies shock morphology and can generate features remarkably similar to the SNRs CAS A and SN 1006 with protrusions in the shell. This study provides a novel perspective for understanding the formation of protrusions in the shells of SNRs.

\section{Simulation Setup}
\label{sec_simu}
Based on the 2D PIC simulation method, we investigate the dynamical evolution process of supernova remnant ejecta in a uniform interstellar medium. The entire calculation is performed using the \textsc{Smilei} code \cite{2018CoPhC.222..351D}.

To simulate the dynamical evolution and particle acceleration during the expansion of an SNR in a uniform interstellar medium, we assume that both the supernova ejecta and the interstellar medium consist of plasma composed of protons and electrons, with electron mass $m_e$ and proton mass $m_p=30\,m_e$. The density of the number of both electrons and protons in the surrounding interstellar medium is denoted by $n_i$. Within a spherical region of radius $r < R_{\text{ini}}$, the supernova ejecta are distributed with a density $n_{\mathrm{s}}$. The temperatures of both the interstellar medium and the ejecta are set to $ T= 2\times10^{-3} m_e c^2$ to approximate a cold plasma. Initially, the ejecta possess a number density of $n_{\mathrm{ej}} = \xi n_{\mathrm{s}}$ for both the electrons and the protons with an outward radial velocity of $v_0$. The interstellar medium contains a magnetic field aligned with the x-axis with $B_x = B_0$, characterized by a magnetization parameter $\sigma$, defined as the
ratio of magnetic energy density to particle kinetic energy density, i.e., $\sigma = B_0^2 / (8 \pi (n_{e^-}m_e  + n_pm_p)v_0^2)$.

The simulation has a spatial scale of $L_x \times L_y = 512  \, c/\omega_{\mathrm{pe}}  \times 512  \, c/\omega_{\mathrm{pe}} $ in the $x$-$y$ plane, where $\omega_{\mathrm{pe}} = \sqrt{4 \pi n_e e^2 / m_e}$ is the electron plasma frequency. The grid resolution is set to $\Delta x = \Delta y = 0.065 c/\omega_{\mathrm{pe}}$, resolving the electron skin depth with high accuracy. Each cell contains $16$ particles per species, and the time step is chosen as $\Delta t = 1/32\ \omega_{\mathrm{pe}}^{-1}$.

\section{Results}\label{sec2}

\begin{figure*}
    \centering
    \includegraphics[width=\textwidth]{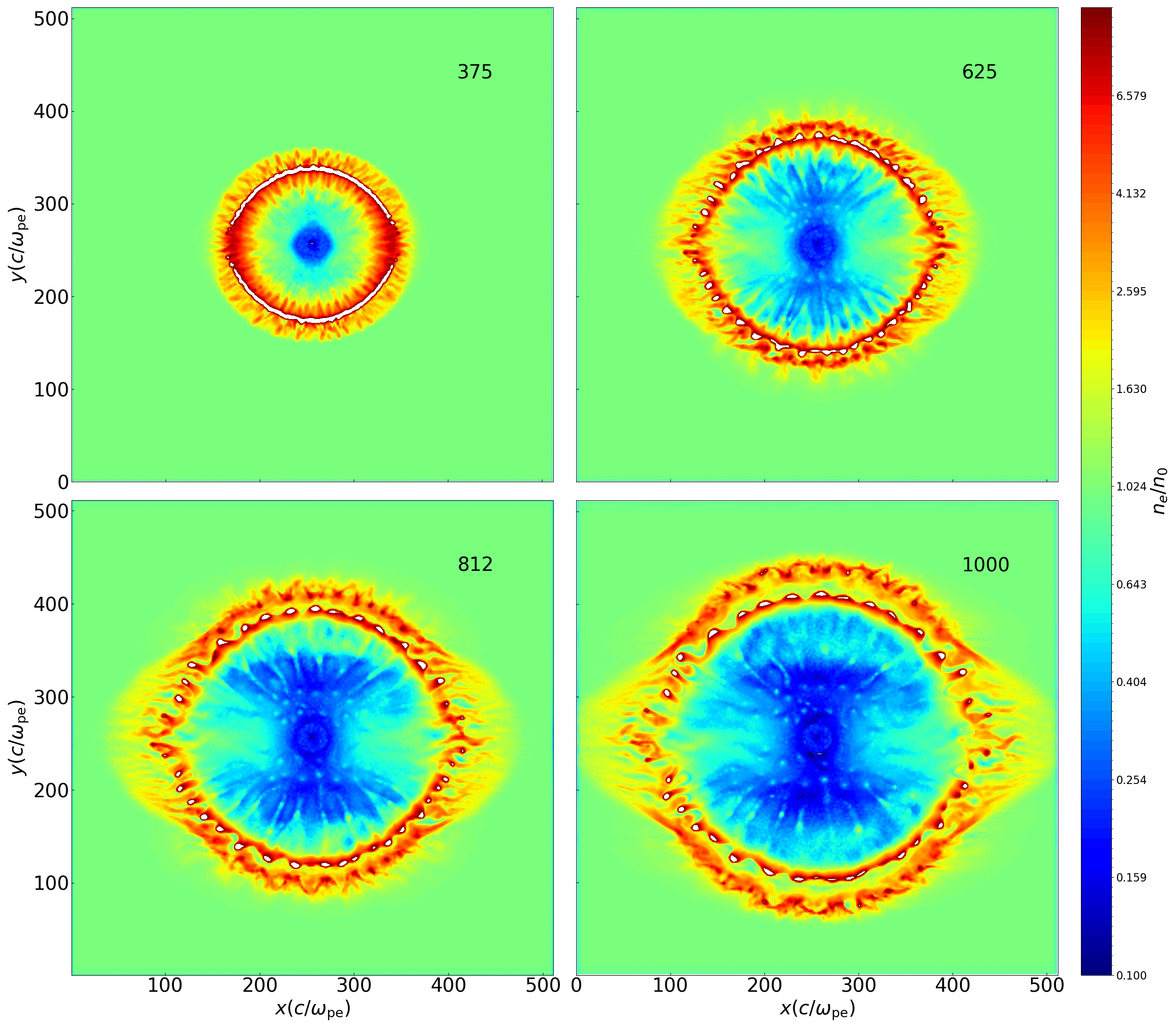}
    \caption{Profiles of the densities of the protons at $t=375 \, \omega_{\mathrm{pe}}^{-1}$ (top left),
    $625 \,\omega_{\mathrm{pe}}^{-1}$ (top right), $812 \, \omega_{\mathrm{pe}}^{-1}$ (bottom left), and
    $1000 \, \omega_{\mathrm{pe}}^{-1}$ (bottom right), respectively,  for the parameters of $\xi = 100$, $R_{\text{ini}} = 16\, c/\omega_{\mathrm{pe}}$, $v_0 = 0.2 \, c$, and $\sigma=10^{-4}$.}
    \label{fig:pdtime}
\end{figure*}

Figure \ref{fig:pdtime} present the evolution of the spatially resolved distributions of  the number densities of protons obtained from the simulations for the parameters of $\xi = 100$, $R_{\text{ini}} = 16\, c/\omega_{\mathrm{pe}}$, $v_0 = 0.2 \, c$, and $\sigma=10^{-4}$. Initially, the high-velocity ejected material expands rapidly into the surrounding interstellar medium, creating a forward shock wave that continuously shocks the interstellar medium. The density and temperature of the shocked medium increase. Simultaneously, a reverse shock wave forms internally, persistently energizing the ejected material. Between the excited interstellar medium and the energized ejected material, a contact discontinuity surface forms, giving rise to the double-shock structure characteristic of SNRs. At the early stage ($t = 375 \, \omega_{\mathrm{pe}}^{-1}$), the forward shock exhibits a largely spherical structure. As the shock expands outward, in the region near $y = 0$, the normal direction of the shock front is approximately aligned with the magnetic field orientation in the upstream medium, forming a quasi-parallel shock. In this configuration, electrons and protons near the shock front can diffuse farther upstream along the magnetic field lines. Conversely, in the shock region near x = 0, the angle between the shock normal and the upstream magnetic field direction is approximately $90^{\circ}$, corresponding to a quasi-perpendicular shock. In these regions, charged particles near the shock front cannot diffuse effectively upstream. Consequently, at later evolutionary stages, the forward shock significantly deviates from a spherical shape, developing a bulging structure in the equatorial region and giving rise to the ear-like features observed in SNRs.

\begin{figure}
    \centering
    \includegraphics[width=0.5\textwidth]{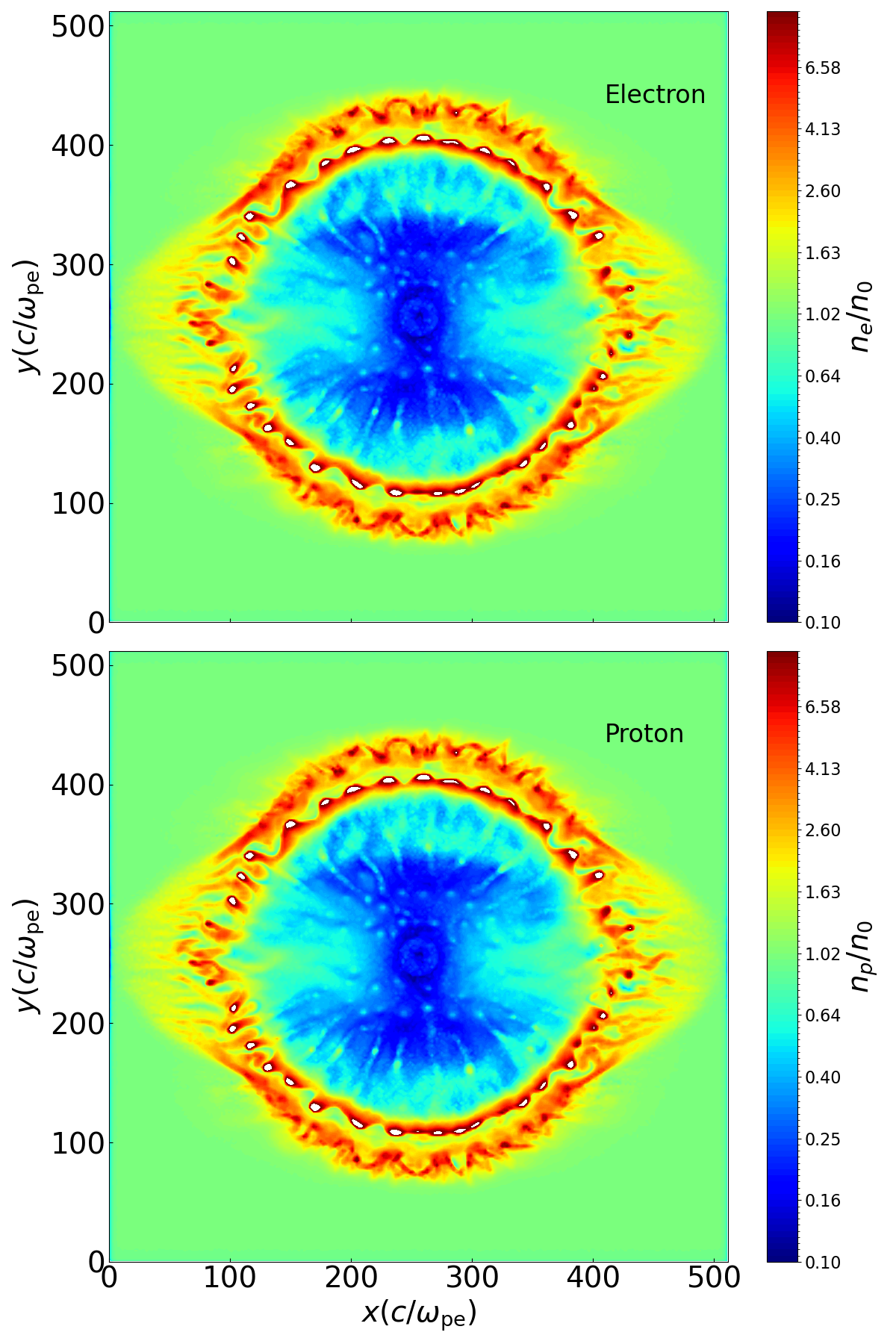}
    \caption{Profiles of the densities of the electrons (top panel) and the protons (bottom panel) at $t=938 \, \omega_{\mathrm{pe}}^{-1}$.}
    \label{fig:density938}
\end{figure}

\begin{figure}
    \centering
    \includegraphics[width=0.5\textwidth]{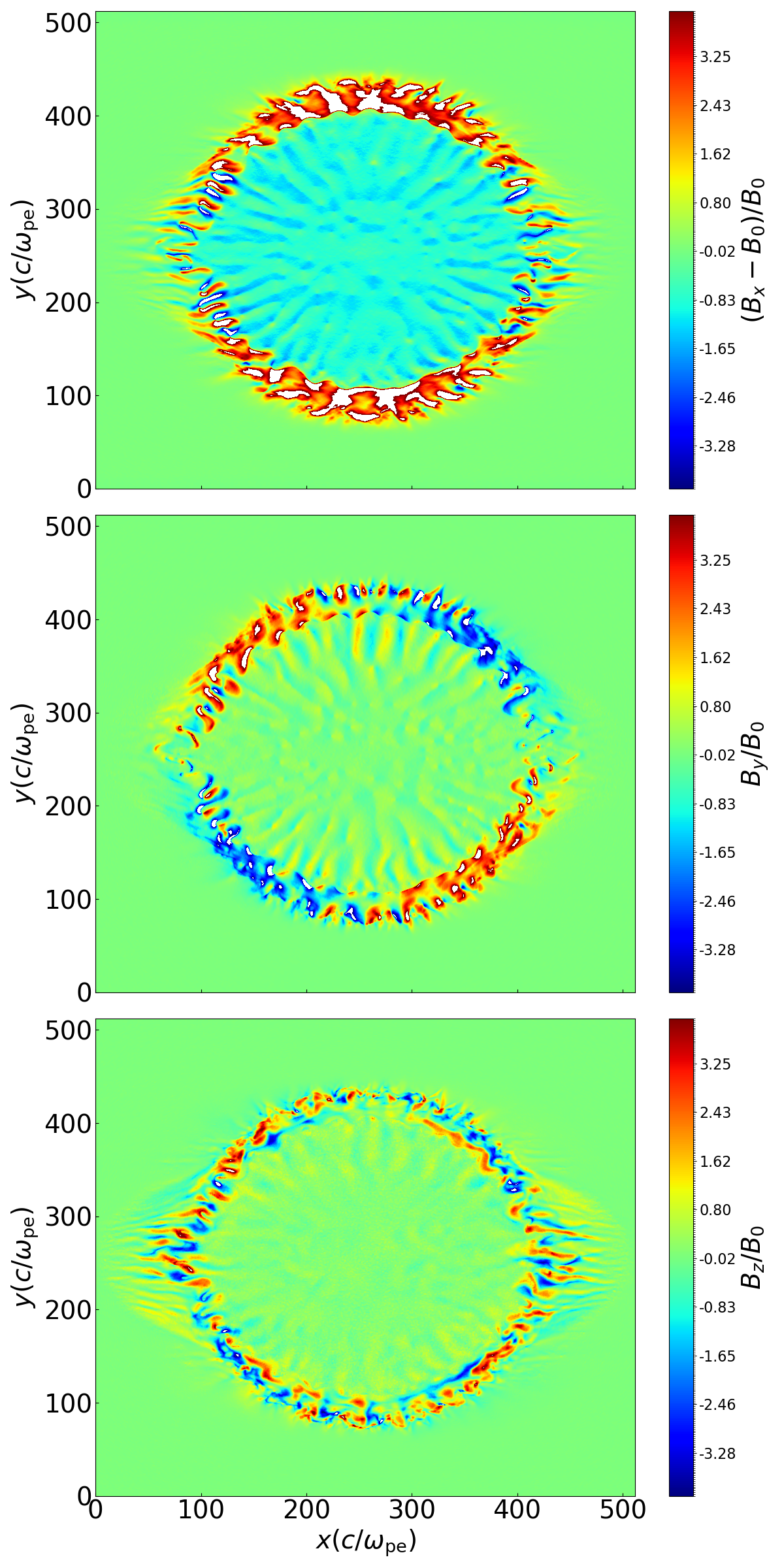}
    \caption{Profiles of the magnetic field components $B_x - B_0$ (top panel), $B_y$ (middle panel) and $B_z$ (bottom panel) at $t=938 \, \omega_{\mathrm{pe}}^{-1}$.}
    \label{fig:mag938}
\end{figure}

At $t=938 \, \omega_{\mathrm{pe}}^{-1}$, the spatial distributions of proton and electron number densities obtained from numerical simulations are shown in Figure \ref{fig:density938}. They are coupled together and exhibit identical spatial distribution characteristics. The spatial distributions of the magnetic field components in the $x$, $y$, and $z$ directions are presented in Figure \ref{fig:mag938}. Within the bulging shell structure near the equatorial region, the distributions of $B_x - B_0$, $B_z$, and particles all display filamentary structures along the x-direction, which is also why particles diffuse more easily along the background magnetic field direction. The shock morphology we obtained is remarkably similar to the ear-like structures observed in the SNRs Simeis 147 \citep{2006A&A...454..239G} and G309.2-00.6 \citep{1998MNRAS.299..812G}.

\begin{figure*}
    \centering
    \parbox{0.48\textwidth}{\includegraphics[width=0.5\textwidth]{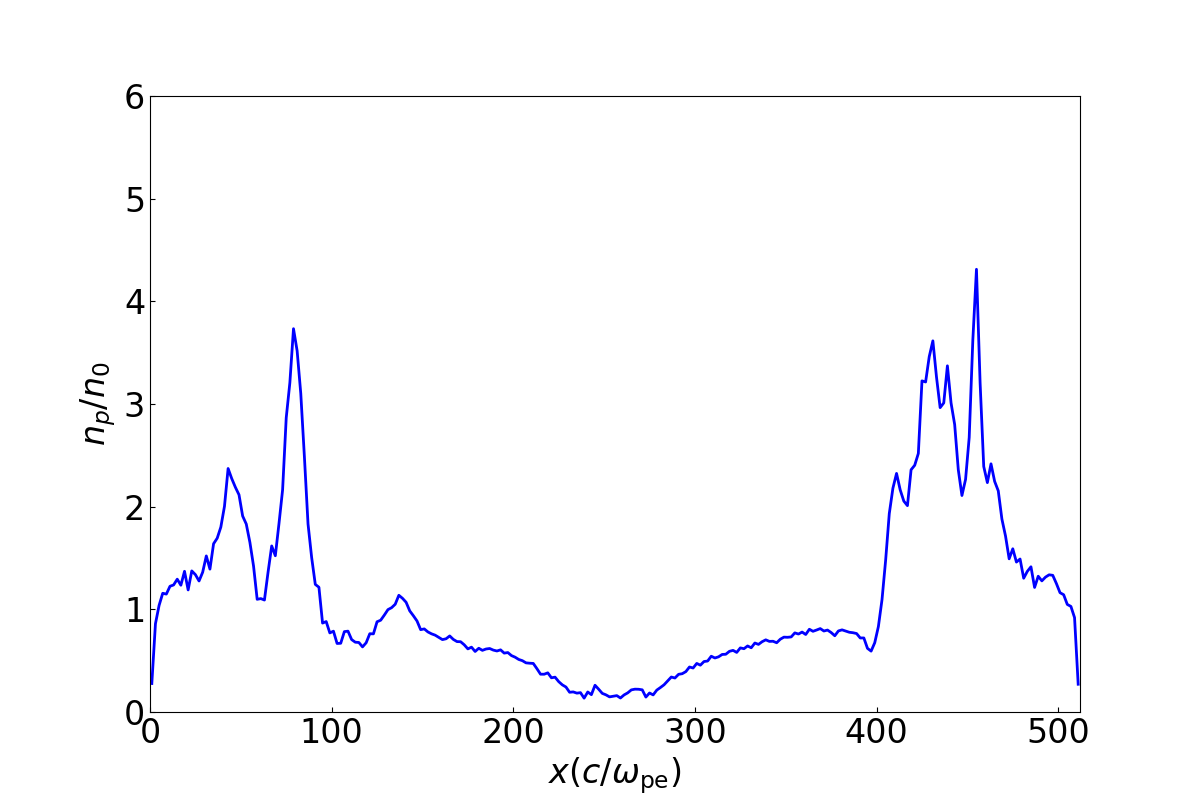}}
    \hfill
    \parbox{0.48\textwidth}{\includegraphics[width=0.5\textwidth]{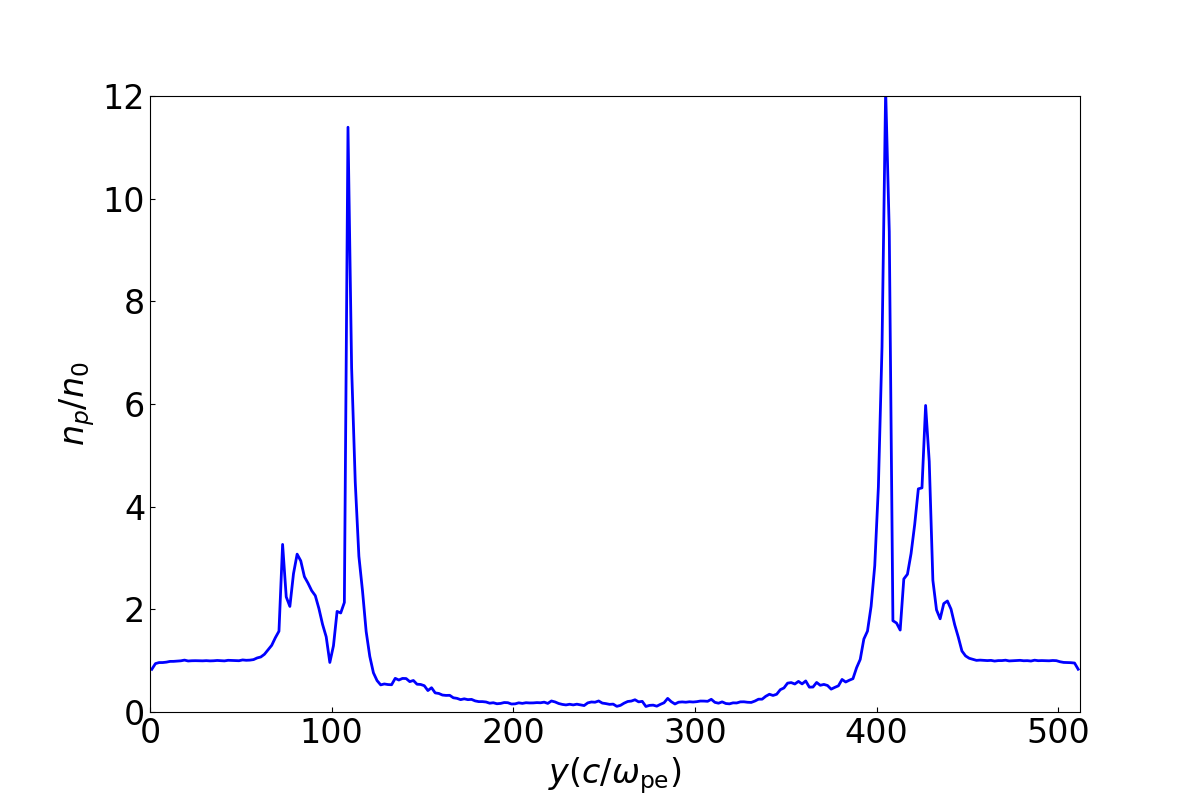}}

    \caption{Proton number density distribution along $y=L_y/2$ (left panel) and $x=L_x/2$ (right panel) at $t=938 \, \omega_{\mathrm{pe}}^{-1}$.} 
    \label{fig:pnx0y0}
\end{figure*}

At $y = L_y/2$, the shock front normal of the forward shock formed around the ejecta is parallel to the upstream magnetic field, classifying it as a parallel shock. Figure \ref{fig:pnx0y0} illustrates the proton number density along $y=L_y/2$  and $x=L_x/2$ at $t=938 \, \omega_{\mathrm{pe}}^{-1}$, when the forward shock sweeps through the upstream medium, it compresses the material. The compression ratio reaches its maxima at $x \sim 50$ and $450$, with another peak at $x \sim 80$ and $420$, corresponding to the locations of the reverse shock. In the regions with $x \leq 50$ and $x \geq 450$, shock precursors form where the medium's compression ratio gradually varies, resulting in a bulging shell structure. On the other hand, along $x = L_x/2$, the normal vector of the forward shock front is perpendicular to the upstream magnetic field direction, constituting a perpendicular shock. In perpendicular shocks, particles near the shock front cannot significantly diffuse into the upstream region, preventing the formation of prominent shock precursors. The forward shock morphology exhibits distinct anisotropy, with the equatorial shock radius exceeding the polar radius  by a factor of $\sim 1.2$ at $t=938 \, \omega_{\mathrm{pe}}^{-1}$.

\section{Conclusions and Discussion}\label{sec12}

Our 2D PIC simulations establish a fundamental connection between magnetic field geometry and the formation of ear-like structures in SNRs. The results demonstrate that particle acceleration preferentially aligned with the ambient magnetic field direction can produce morphological features strikingly similar to observed SNR protrusions, without invoking jets or circumstellar medium asymmetries. This mechanism provides an alternative explanation for the origin of these enigmatic structures, and it is complementary to the existing jet-based and circumstellar interaction models.

The simulations reveal a clear dependence of shock morphology on the orientation of the upstream magnetic field. Along the $y = L_y/2$ plane where the shock normal parallels the upstream magnetic field, efficient particle diffusion generates extended protrusions, as evidenced in Figure \ref{fig:pnx0y0} (left panel). Conversely, at $x = L_x/2$ where the shock becomes quasi-perpendicular, restricted particle transport maintains a compact shock front (Figure \ref{fig:pnx0y0}, right panel). This dichotomy explains the development of axisymmetric ears from initially spherical ejecta.


These findings bear significant resemblance to observational characteristics of well-studied SNRs. The simulated ears show particular morphological affinity with the northeast and southwest limbs of SN 1006, where radio polarization studies have identified quasi-parallel shock geometries \citep{2013AJ....145..104R}.
Similarly, the bipolar protrusions in Cassiopeia A may reflect analogous magnetic alignment effects rather than exclusively jet-driven phenomena. While jet scenarios \citep{1998ApJ...493..781G} and CSM interactions \citep{2021MNRAS.502..176C} remain viable for certain systems, our results demonstrate that magnetic field orientation and particle acceleration can produce comparable structures through different physical processes.

Recent work by \citet{2025A&A...694A.202B} proposed that ear-like structures in Type Ia SNRs may arise from either anisotropic cosmic-ray acceleration (via spatially varying $\gamma_{\text{eff}}$) or ISM density gradients, with hydrodynamic simulations reproducing protrusions akin to Kepler's SNR and G1.9+0.3. Their findings complement our PIC results, where magnetic field-aligned particle acceleration drives ear formation, but differ in the predicted shock morphology: while their $\gamma_{\text{eff}}$ scenario produces arched bulges from ``shrinking'' intershock regions, our model links protrusions directly to quasi-parallel shock geometry and field-guided diffusion. We elucidate the underlying kinetic origin of the asymmetry by directly linking protrusion formation to efficient particle diffusion in quasi-parallel shock geometries

Future investigations should address several important extensions of this work. Three-dimensional simulations incorporating oblique magnetic field orientations would better approximate realistic astrophysical conditions. Coupling our PIC-scale results with global MHD models could bridge the gap between kinetic processes and remnant-scale evolution. Most crucially, generating synthetic non-thermal emission maps would enable direct quantitative comparison with multi-wavelength observations of SN 1006 and similar remnants.

This study expands our understanding of SNR morphology by demonstrating how magnetic field-guided particle acceleration can generate characteristic ear-like structures. The mechanism provides a unified framework for interpreting asymmetric features across different remnant classes, while offering observable predictions to distinguish magnetic alignment from jet or CSM interaction scenarios.

\section*{Data availability}
The simulations in this paper were generated using the publicly code \textsc{Smilei}. The data underlying this article will be shared on reasonable request to the corresponding author.

\section*{Acknowledgements}

This work was supported by the National Natural Science
Foundation of China (grant Nos. 12393852 and 12563010), Yunnan Fundamental Research Projects (grant No. 202501AS070068),  and Yunnan Revitalization Talent Support Program (XDYC-QNRC-2022-0486).
\bibliographystyle{mnras}
\bibliography{reference}

\begin{thebibliography}{}
\expandafter\ifx\csname natexlab\endcsname\relax\def\natexlab#1{#1}\fi
\providecommand{\url}[1]{\href{#1}{#1}}
\providecommand{\dodoi}[1]{doi:~\href{http://doi.org/#1}{\nolinkurl{#1}}}
\providecommand{\doeprint}[1]{\href{http://ascl.net/#1}{\nolinkurl{http://ascl.net/#1}}}
\providecommand{\doarXiv}[1]{\href{https://arxiv.org/abs/#1}{\nolinkurl{https://arxiv.org/abs/#1}}}

\bibitem[\protect\citeauthoryear{Bear, Grichener, \& Soker}{2017}]{2017MNRAS.472.1770B}
Bear E., Grichener A., Soker N., 2017, MNRAS, 472, 1770. doi:10.1093/mnras/stx2125

\bibitem[\protect\citeauthoryear{Bamba et al.}{2003}]{2003ApJ...589..827B} Bamba A., Yamazaki R., Ueno M., Koyama K., 2003,
ApJ, 589, 827. doi:10.1086/374687

\bibitem[\protect\citeauthoryear{Bao et al.}{2025}]{2025A&A...694A.202B} Bao B., Wang Y., Yang C., Zhang L., 2025,
A\&A, 694, A202. doi:10.1051/0004-6361/202449381

\bibitem[\protect\citeauthoryear{Chiotellis, Boumis, \& Spetsieri}{2020}]{2020Galax...8...38C}
Chiotellis A., Boumis P., Spetsieri Z.~T., 2020, Galax, 8, 38. doi:10.3390/galaxies8020038

\bibitem[\protect\citeauthoryear{Chiotellis, Boumis, \& Spetsieri}{2021}]{2021MNRAS.502..176C} Chiotellis A., Boumis P.,
Spetsieri Z.~T., 2021, MNRAS, 502, 176. doi:10.1093/mnras/staa3573

\bibitem[\protect\citeauthoryear{Gaensler, Green, \& Manchester}{1998}]{1998MNRAS.299..812G}
Gaensler B.~M., Green A.~J., Manchester R.~N., 1998, MNRAS, 299, 812. doi:10.1046/j.1365-8711.1998.01814.x

\bibitem[\protect\citeauthoryear{Derouillat et al.}{2018}]{2018CoPhC.222..351D} Derouillat J., Beck A., P{\'e}rez F., Vinci T., Chiaramello M.,
Grassi A., Fl{\'e} M., et al., 2018, CoPhC, 222, 351. doi:10.1016/j.cpc.2017.09.024


\bibitem[\protect\citeauthoryear{Fang et al.}{2020}]{2020MNRAS.491.2460F} Fang J., Yan J., Wen L., Lu C., Yu H., 2020,
MNRAS, 491, 2460. doi:10.1093/mnras/stz3214

\bibitem[\protect\citeauthoryear{Gaensler}{1998}]{1998ApJ...493..781G} Gaensler B.~M., 1998, ApJ, 493, 781. doi:10.1086/305146

\bibitem[\protect\citeauthoryear{Grichener \& Soker}{2017}]{2017MNRAS.468.1226G} Grichener A.,
Soker N., 2017, MNRAS, 468, 1226. doi:10.1093/mnras/stx534

\bibitem[\protect\citeauthoryear{Gvaramadze}{2006}]{2006A&A...454..239G}
Gvaramadze V.~V., 2006, A\&A, 454, 239. doi:10.1051/0004-6361:20054114

\bibitem[\protect\citeauthoryear{Li et al.}{2015}]{2015MNRAS.453.3953L} Li J.-T.,
Decourchelle A., Miceli M., Vink J., Bocchino F., 2015, MNRAS, 453, 3953. doi:10.1093/mnras/stv1882

\bibitem[\protect\citeauthoryear{Miceli et al.}{2009}]{Miceli2009} Miceli M., Bocchino F., Iakubovskyi D., Orlando S.,
Telezhinsky I., Kirsch M.~G.~F., Petruk O., et al., 2009, A\&A, 501, 239. doi:10.1051/0004-6361/200811505

\bibitem[\protect\citeauthoryear{Reynoso, Hughes, \& Moffett}{2013}]{2013AJ....145..104R} Reynoso E.~M., Hughes J.~P., Moffett D.~A.,
2013, AJ, 145, 104. doi:10.1088/0004-6256/145/4/104

\bibitem[\protect\citeauthoryear{Reynolds \& Gilmore}{1986}]{1986AJ.....92.1138R} Reynolds S.~P., Gilmore D.~M., 1986, AJ, 92, 1138. doi:10.1086/114244

\bibitem[\protect\citeauthoryear{Rothenflug et al.}{2004}]{2004A&A...425..121R} Rothenflug R., Ballet J., Dubner G., Giacani E.,
Decourchelle A., Ferrando P., 2004, A\&A, 425, 121. doi:10.1051/0004-6361:20047104

\bibitem[\protect\citeauthoryear{Winkler et al.}{2014}]{2014ApJ...781...65W} Winkler P.~F., Williams B.~J., Reynolds S.~P.,
Petre R., Long K.~S., Katsuda S., Hwang U., 2014, ApJ, 781, 65. doi:10.1088/0004-637X/781/2/65
\end{thebibliography}

\bsp
\label{lastpage}
\end{document}